# 5GNOW: Challenging the LTE Design Paradigms of Orthogonality and Synchronicity

Gerhard Wunder[1], Martin Kasparick[1], Stephan ten Brink[2], Frank Schaich[2], Thorsten Wild[2], Ivan Gaspar[3], Eckhard Ohlmer[3], Stefan Krone[3], Nicola Michailow[3], Ainoa Navarro[3], Gerhard Fettweis[3], Dimitri Ktenas[4], Vincent Berg[4], Marcin Dryjanski[5], Slawomir Pietrzyk[5], Bertalan Eged[6]

[1]Fraunhofer Heinrich Hertz Institut, Berlin, Germany
[2]Bell Labs, Alcatel-Lucent, Stuttgart, Germany
[3]TUD (Technische Universität Dresden) – Vodafone Chair Mobile Communications Systems, Dresden, Germany
[4]Commissariat à l'énergie atomique et aux énergies alternatives (CEA – Leti), MINATEC Campus, Grenoble, France
[5]IS-Wireless, Warsaw, Poland
[6]Systems Engineering Center of Excellence, National Instruments, Budapest, Hungary
*www.5gnow.eu*

*Abstract*— LTE and LTE-Advanced have been optimized to deliver high bandwidth pipes to wireless users. The transport mechanisms have been tailored to maximize single cell performance by enforcing strict synchronism and orthogonality within a single cell and within a single contiguous frequency band. Various emerging trends reveal major shortcomings of those design criteria:
• The fraction of machine-type-communications (MTC) is growing fast. Transmissions of this kind are suffering from the bulky procedures necessary to ensure strict synchronism.
• Collaborative schemes have been introduced to boost capacity and coverage (CoMP), and wireless networks are becoming more and more heterogeneous following the non-uniform distribution of users. Tremendous efforts must be spent to collect the gains and to manage such systems under the premise of strict synchronism and orthogonality.
• The advent of the Digital Agenda and the introduction of carrier aggregation are forcing the transmission systems to deal with fragmented spectrum.
5GNOW (5th Generation Non-Orthogonal Waveforms for Asynchronous Signalling) is an European collaborative research project supported by the European Commission within FP7 ICT Call 8. 5GNOW will question the design targets of LTE and LTE-Advanced having these shortcomings in mind. The obedience of LTE and LTE-Advanced to strict synchronism and orthogonality will be challenged. It will develop new PHY and MAC layer concepts being better suited to meet the upcoming needs with respect to service variety and heterogeneous transmission setups. A demonstrator will be built as Proof-of-Concept relying upon continuously growing capabilities of silicon based processing. Wireless transmission networks following the outcomes of 5GNOW will be better suited to meet the manifoldness of services, device classes and transmission setups being present in envisioned future scenarios like smart cities. The integration of systems relying heavily on MTC, e.g. sensor networks, into the communication network will be eased. The per-user experience will be more uniform and satisfying. To ensure this 5GNOW will contribute to upcoming 5G standardization.

## I. KILLER APPLICATIONS: THE 5G DRIVERS

The successful deployment of killer applications in wireless communication technology has allowed its rapid development in the past 20 years with major impact on modern life and on the way the societies operate in politics, economy, education, entertainment, logistics & travel, and industry.

First and foremost the need for un-tethered telephony and therefore wireless real-time communication has dominated the success of cordless phones, followed by first generation (1G) of cellular communications. Soon, incorporated in 2G, two-way paging implemented by SMS text messaging became the second killer application. With the success of wireless LAN technology (i.e. IEEE 802.11), http internet browsing, and the widespread market adoption of laptop computers internet data connectivity became interesting for anyone, opening up the opportunity for creating a market for the third killer application in 3G, wireless data connectivity. The logical next step has been the shrinkage of the laptop, merging it with the cellular telephone into todays' smartphones, and offering high bandwidth access to wireless users with the world's information at their fingertips everywhere and everytime. This is the scenario of the current 4G generation with the most prominent example LTE-A (Long Term Evolution - Advanced). Hence, smartphones are, undoubtedly, in the focus of service architectures for future mobile access networks. Current market trends and future projections indicate that smartphone sales will keep growing and overtook conventional phones [TIA's 2009/2010/2011/2012 ICT Market Review and Forecast] to constitute now the lion's share of the global phone market: the smartphone has become a mass market device.

The next foreseen killer application is the massive wireless connectivity of machines with other machines, referred to as M2M or the Internet of Things (IoT). During the past years a multitude of wireless M2M applications has been explored, e.g. information dissemination in public transport systems or in manufacturing plants. However, fast deployment of M2M through a simple 'plug and play' connection via cellular networks is not a reality and the commercial success has been somewhat limited, yet. The availability of cellular coverage needs to be combined with simplicity of handling, in both software and hardware aspects, i.e. avoiding having to setup and connect as in a ZigBee or WLAN hot-spot but at the same time allowing longer battery life time and cheap devices. These principles can stimulate subscribers to buy M2M sensors and participate in the collection of monitoring data. M2M can be employed by communities (social network) to share monitoring information about cars, homes and environment, which could lead to a number of connected devices orders of magnitude higher than today.

## II. APPLICATION CHALLENGES

### A. Service differentiation and Gigabit wireless connectivity

The typical use of a smartphone goes beyond simple voice calls as seen in Figure 1. The variety of services ask for covering a much wider range, i.e., bandwidth-hungry applications such as video streaming, latency-sensitive applications such as networked online gaming, and in particular sporadic machine-type-like applications due to smartphone apps which are most of the time inactive but regularly access the internet for minor/incremental updates with no human interaction (e.g. weather forecasts, stock prices, navigation position, location-dependent context information etc.). Hence, the operators have to be well prepared to take on the challenges of a much higher per-user rate, increasing overall required bandwidth and service differentiation threatening the common value chains on which they rely to compensate for investment costs of future user services [1]. From a technical perspective it seems to be utmost challenging to provide uniform service experience to users under the premises of heterogeneous networks.

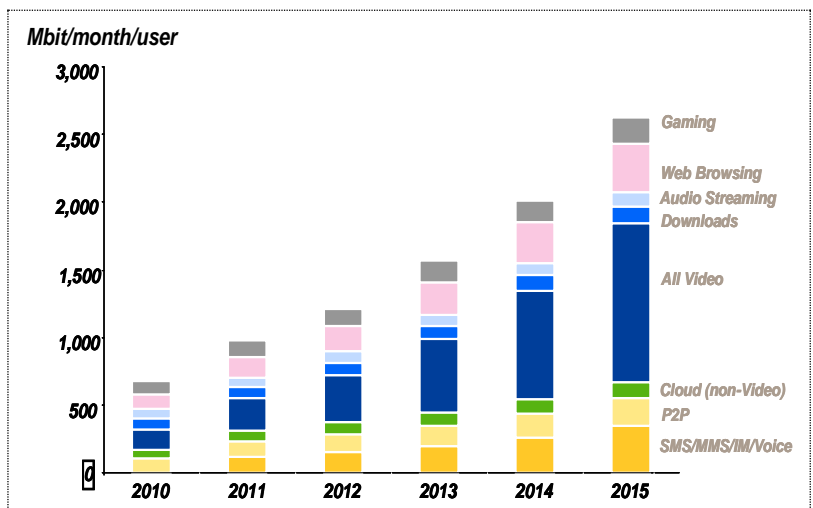

Figure 1: Mixture of traffic types (internal study by ALUD)

It can be also seen in Figure 1, that more than 50% of the data volume measured in cellular networks is already generated by users consuming streaming applications. As high definition 3D streaming with user enabled vision angle control requires on the order of 100Mb/s, and users want quick downloads of typically above 100x real-time of multiple streams, it is possible to see 10-100 Gb/s wireless connectivity coming up as a requirement in future. Obviously this does not lead to a need for a continuously sustainable very high bandwidth for one user over long periods of time. Instead, 100Gb/s data rates will be shared via the wireless medium.

### B. Fast dormancy and sporadic access

Somewhat surprisingly, sporadic access poses another significant challenge to mobile access networks due to an operation known as *fast dormancy* [2][3]. Fast dormancy is used by handheld manufacturers to save battery power by using the feature that a mobile can break ties to the network individually and as soon as a data piece is delivered the smartphone changes from active into idle state. Consequently, when the mobile has to deliver more pieces of data it will always go through the complete synchronization procedure again. Actually, this can happen several hundred times a day resulting in significant control signaling growth and network congestion threat, see Figure 2. Furthermore, with M2M on the horizon, a multitude of (potentially billions) machine-type-communication (MTC) devices accessing asynchronously the network will dramatically amplify the problem.

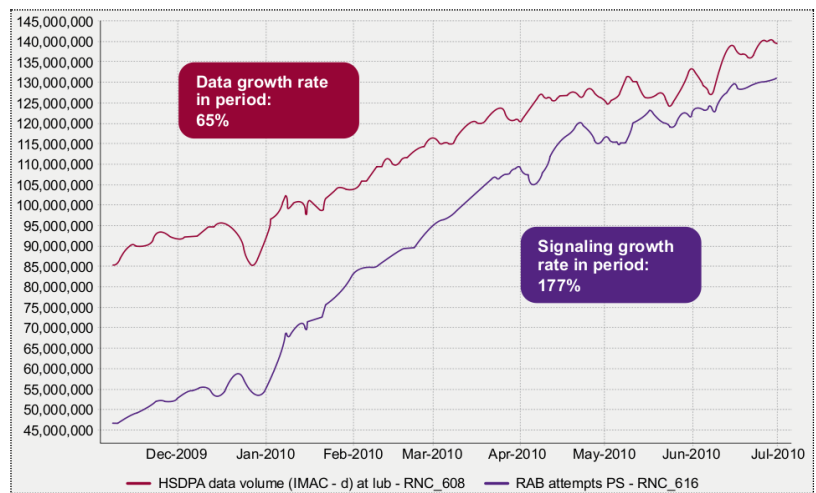

**Figure 2: Comparisons of data and signalling traffic [2]**

### C. Fragmented Spectrum and Carrier aggregation

A third challenge for cellular networks is the variable usage of aggregated non-contiguous frequency bands, so-called carrier aggregation which is implemented to achieve much higher rates. Carrier aggregation implies the use of separate RF front ends accessing different channels and can provide higher data rates in the downlink, reinforcing the attraction of isolated frequency bands such as the L-Band. Actually, the search for new spectrum is very active in Europe and in the USA in order to provide mobile broadband expansion. It includes the opportunistic use of spectrum, which has been an interesting research area in wireless communications in the past decade.

Techniques to detect and assess channel vacancy using cognitive radio (CR) could well make new business models possible in the future. The first real implementation will start with the exploration of TV white spaces (TVWS) in USA. Combined with the preparation of the regulatory framework in Europe, TVWS exploration can represent a new niche market if it overcomes, with spectrum agility, the rigorous implementation requirements of low out of band radiation for protection of legacy systems. In [5] this scenario is detailed and Figure 3 exhibits the loss of efficiency of traditional OFDM with cyclic prefix (CP) to fit in an ESM (Emission Spectrum Mask) due its non-negligible side lobes. The picture helps to see how the partial use of the available spectrum can alleviate the need of additional filtering process, which is in the analog case also represents less insertion loss on the output of the transmitter.

### III. STATUS AND DEVELOPMENTS OF LTE

### A. Toward higher spectral efficiency

LTE Release 8 (R8) and Release 9 (R9) ideally offer data rates of up to 300Mb/s in the downlink and 75Mb/s in the uplink. The push for higher data rates in the subsequent standard update R10 is mostly addressed by going from 4 to 8 transmit antennas in the downlink, and from 1 to 4 transmit antennas in the uplink. R10 is thus compliant to the IMT-Advanced requirements and is known as LTE-Advanced (LTE-A). Even higher data rates (over 1Gb/s in downlink, and over 500Mb/s in uplink) will be possibly achieved through carrier aggregation. Inter-cell coordination like ICIC (inter-cell interference coordination, R10) as well as joint processing enabled by CoMP (Coordinated Multipoint) transmission and reception (R11) will further increase spectral efficiency. All such schemes seek for an evolutionary approach within a synchronous/orthogonal physical layer framework. In 10 years, following the progression of wireless throughputs, rates in the order of 10Gb/s must be addressed.

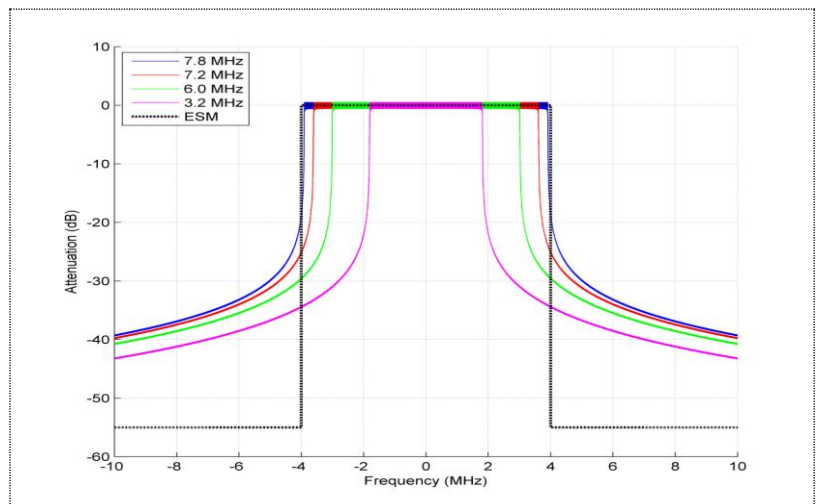

**Figure 3: OFDM+CP vs. ESM, original picture [5]**

*B. Incorporating MTC traffic*

Specific traffic needs of access devices such as smartphones, multimedia tablets or MTC devices have not been addressed in LTE R10 yet. Moreover, particularly MTC devices are currently viewed as a threat to network stability due to signalling overhead, and thus, restrictions are considered upon bringing in such devices into the network (network overload protection). MTC devices may have other special requirements such as low power operation, low data transmission, bursty and sporadic traffic profiles. Therefore a dedicated study group has been formed within 3GPP to which plans to work out signalling and protocol schemes to include MTC traffic with current and future standard update (starting from R11 [4]) which enables creation of multimedia sensor networks at the service level. This contribution could be realized in terms of proposing specific definition of services, requirements, features, options, signalling, and many more. MTC, as a specific type of communication, has its special requirements, such as for example a strong need to minimize the power consumption at the end device. Other example features of MTC to be proposed and evaluated include mobile-only originated calls, low data transmission, low power transmission, low mobility scenarios (e.g. surveillance cameras), periodic data transmission, and delay tolerant services. Also here, a fundamental issue for working on MTC is to have an adequate and detailed model of the PHY and MAC layers of the system. So far only the basic features and concept of MTC are included [4] at the level of stage 1 specification. Physical layer techniques, however, are not within the scope of such activities. Some MAC amendments may be considered. Both offer a great opportunity to exceed (PHY) or contribute (MAC) to the specifications for R12 and beyond.

*C. Fragmented spectrum*

LTE is, one way or another, dealing with some spectrum agility as a requisite to allow worldwide interoperability of devices in a fragmented spectrum, fuelled by ongoing spectrum auctions, license renewals and re-farming initiatives across a wide range of frequency band [8]. LTE is already implemented in many frequency bands: 700 MHz and AWS (Advanced Wireless Spectrum) bands in the USA, 800, 1800 and 2600 MHz in Europe, 2100 MHz and 2600 MHz in Asia. It is expected that the 2600 MHz, 1800 MHz and 800 MHz bands are the most widely used in Western Europe for 4G deployments.

As a conclusion, the lack of spectrum harmonization is forcing vendors to find contour solutions to deliver, as far as possible, globally compatible LTE chipsets and devices (and OFDM spectrum property is not helping much to alleviate the analog signal processing challenges in the front-end).

IV. 5GNOW APPROACH

5GNOW (5th Generation Non-Orthogonal Waveforms for Asynchronous Signaling) is a European collaborative research project supported by the European Commission within FP7 ICT Call 8 (09/2012-02/2015). The scope of the project is described next.

*A. Questioning the common design priciples*

*The main hypothesis of this explorative research project is that, specifically, the underlying design principles –synchronism and orthogonality– of the PHY layer of today's mobile radio systems constitute a major obstacle for the envisioned service architecture. Hence, there is a clear motivation for an innovative and in part disruptive re-design of the PHY layer from scratch.*

5GNOW explores this key hypothesis along the application scenarios MTC, CoMP/HetNet, and carrier aggregation. For example, MTC traffic generating devices (including smartphones) should not be forced to be integrated into the bulky synchronization procedure which has been deliberately designed to meet orthogonal constraints. Instead, they optimally should be able to awake only occasionally and transmit their message right away only coarsely synchronized. By doing so MTC traffic would be removed from standard uplink data pipes with drastically reduced signalling overhead. Therefore, alleviating the synchronism requirements can significantly improve operational capabilities and network performance as well as user experience and life time of smartphones and autonomous MTC nodes. 5GNOW will follow up on this idea and develop non-orthogonal waveforms for asynchronous signalling in the uplink (and specifically RACH) to enable such efficient MTC traffic.

MTC traffic and the corresponding network congestion problems are primarily concerned with synchronism and orthogonality constraints on the LTE-A PHY layer uplink channels. However, LTE-A downlink is also involved when Coordinated Multipoint (CoMP) transmission [6] is considered. CoMP is driven by the appealing idea to exploit the superior single-cell performance of the underlying synchronous orthogonal air interfaces enabling both uniform coverage and high capacity. However, such an approach entails huge additional overhead in terms of message sharing, base station synchronization, feedback of channel state information, forwarding of control information etc. On top, the approach is known to lack robustness against the actual extent to what the delivered information reflects the current network state - in fact, it turns out that the achieved gains by CoMP transmission are still far away from the theoretical limits while even constraining the potential services in the network due to extensive uplink capacity use for control signalling [9]. In addition, in a heterogeneous networking scenario (HetNet) with uncoordinated pico or femto cells and highly overlapping coverage, as in today's networks, it seems illusive to provide the required information to all network entities. Evidently, if the degree of coordination to maintain synchronism and orthogonality across layers (PHY, MAC/networking) is not attainable the LTE-A PHY layer should not be forced into such strict requirements, calling for 5GNOW non-orthogonal waveforms for asynchronous signalling and also improved robustness in the downlink.

Finally, interference within the network is overlaid by certainly likewise inherently uncoordinated interference from other legacy networks due to carrier aggregation. Current systems impose generous guard bands to satisfy spectral mask requirements which either severely deteriorate spectral efficiency or even prevent band usage at all, which is again an artefact of strict orthogonality and synchronism constraints within the PHY layer. 5GNOW will address carrier aggregation by implementing sharp frequency notches using non-orthogonal waveforms in order not to interfere with other legacy systems and tight spectral masks.

*B. Enabler for new PHY*

To enable the 5GNOW approach the basic concept of this project is to dismiss the widely unquestioned credo of strict synchronism and orthogonality in the network and, instead, to introduce a broader non-orthogonal robustness concept incorporating the overall required control signalling effort and the applied waveforms in a joint framework. At the core of this paradigm is the introduction of new non-orthogonal waveforms that carry the data on the physical layer. The idea is to abandon synchronism and orthogonality altogether, thereby admitting some crosstalk or interference, and to control these impairments by a suitable transceiver structure and transmission technique. The technological challenges are manifold and require advanced and, most likely, more complex transceiver designs. Due to the evolving silicon processing capabilities, according to Moore's law, today's receiver complexity is about one order of magnitude less constrained than it was several years ago, when LTE Release 8 was being designed. Current LTE baseband signal processing, according to Figure 4, consumes less than 20% of the overall mobile phone power budget, with the 'inner receiver', consisting of signal detection and parameter estimation, consuming less than 10% of the overall baseband operations. With a boost from Moore's law, it is self-evident that 5G smartphones will have plenty of headroom for inner receiver complexity increases, compared to 3.5G, as needed for processing non-orthogonal, asynchronous signals.

5GNOW investigates the inherent tradeoffs between possible relaxation in orthogonality and synchronism and their corresponding impact on performance and network operation/user experience versus the required signal processing capabilities. The research project will make use of the natural emerging technological possibilities suitable for 5G communication. All partners will use their previous experience from former projects (e.g. EU FP7 PHYDYAS) to address the non-orthogonal, asynchronous approach within the trinity of the corner scenarios MTC, CoMP/HetNet, carrier aggregation.

It is important to note that the followed approach stands in clear contrast to (potentially competing) approaches where the challenges are met by applying higher bandwidths and/or increasing the number of antennas. In such approaches the network cost of establishing and maintaining orthogonality and synchronism is mostly overlooked devouring much of the promised gains. A more comprehensive system view is necessary, though, to enable more scalable network architectures.

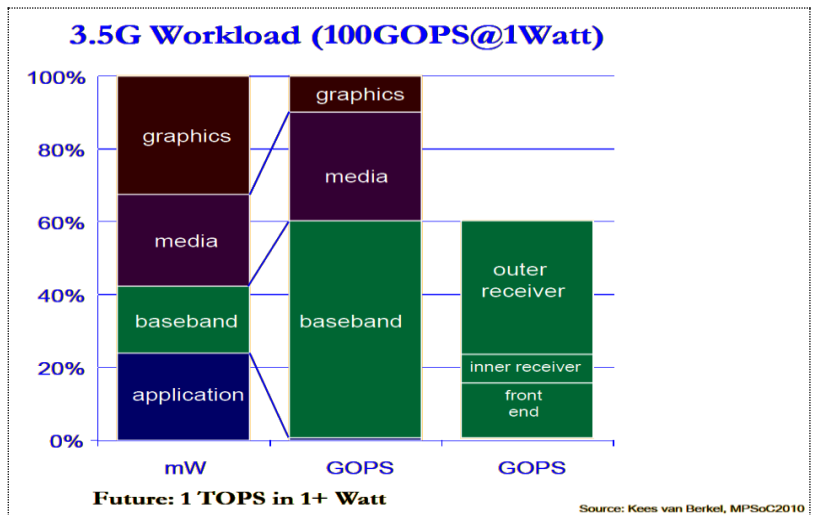
**Figure 4: Workload of current mobiles (outer receiver consists of channel decoder and de-interleaver)**

V. NOVEL IDEAS AND KEY RESEARCH AREAS

The conceptual paradigm shift from synchronous and orthogonal to asynchronous and non-orthogonal systems with increased robustness pose a number of research challenges which shall be addressed in this project. In the center is the design of a new PHY layer for 5G systems using non-orthogonal waveforms which can alleviate the synchronism requirements of existing systems and support all the service requirements for future mobile access networks. These advanced techniques shall be devised in this project and will be addressed through the following main research areas.

*A. Physical Layer*

5GNOW addresses the design of a 5G PHY layer for asynchronous signalling and increased robustness in point-to-multipoint (or conversely) multiple-input-multiple-output (MIMO) transmission using non-orthogonal waveforms. The research will be investigated along the key scenarios MTC, CoMP/HetNet, carrier aggregation. Specifically, the novel key ideas and corresponding scientific and technological (S/T) objectives on PHY layer are:

1. Design of a non-orthogonal PHY layer RACH channel to enable asynchronous MTC traffic by using state-of-the-art sparse MIMO signal processing methodology at the base station. The recognisability and measurability of this objective is high because the integration of MTC communication into the network is highly simplified. This makes the system more capable to serve uncoordinated sporadic access of many nodes and to handle unforeseen events when many nodes abruptly access the network concurrently e.g. in case of emergency situations. This will be measurable by a reduced congestion probability within the heterogeneous scenarios and other appropriate measures to be developed, e.g. such as life time of MTC nodes, infrastructure costs etc.

2. Provision of CoMP/HetNet concepts based on non-orthogonal MIMO PHY layer and related transceiver design and transmission techniques in the presence of relaxed time/frequency synchronization requirements and imperfect channel state information. Some initial waveforms candidate list are e.g. Generalized Frequency Division Multiplexing (GFDM) [10]-[14], Filter Bank Multicarrier (FBMC) [14][15].
The objective is measured in terms of reduced signalling effort and backhaul traffic while maintaining some benchmark performance as well as scalability of the whole system in terms of e.g. rate, waveform properties (e.g. peak-to-average power ratio), out-of-band radiation, ability to seamlessly accommodate many sporadic, low rate users and high rate users; the objective also entails the definition of indicators measuring the allowable degree of asynchronism by non-orthogonal system design and extraction and abstraction of physical layer parameters to be used by upper layers for system operation.

3. Provision of carrier aggregation/fragmented spectrum transmission techniques using non-orthogonal waveforms with relaxed (or completely omitted) synchronization requirements. The objective will be measured in terms of increased bandwidth efficiency in white spaces.

4. Scenario-dependent (channel profile etc.) waveform adaption along key scenarios described before for seamless system operation. The objective is measured in terms of relative gains to a static setup.

*B. MAC layer*

5GNOW addresses also the adaption of the respective system aspects to enable efficient and scalable multi-cell operation within heterogeneous environment. This specifically implies a re-design of (selected) MAC/networking key functionalities for the underlying new PHY layer targeting MAC-related impairments. The novel key ideas and corresponding S/T objectives on MAC layer are:

1. The developed non-orthogonal PHY layer waveforms and their specific structure will be incorporated into the design of the control signalling on MAC layer leading to different designs for different waveforms. Such approach (see e.g. [7] incorporating spatial transmit codebooks) differs significantly from state-of-the-art methodology where typically the transmitted signals have no impact on the design of the control channel. The objective is measured in terms of reduced control signalling and appropriate metrics for the system's sensitivity to mobility, capacity limitations etc.
2. Provision of non-orthogonal scheduling framework measurable in terms of scheduler performance gains along sum throughput and standard fairness metrics.

*C. Demonstrator*

5GNOW provides a proof-of-concept for the non-orthogonal physical layer techniques/algorithms by means of a hardware demonstrator and over-the-air transmission experiments. The feasibility and potential benefits of selected 5G core concepts shall be demonstrated using a hardware platform. Hence, reality check of the proposed concepts via lab demonstrations is a measurable objective.

The performance will be compared to a standard LTE-A air interface using suitable performance indicators such as transmission rate, the achieved relaxation of synchronism and orthogonality requirements and its impact on control signalling overhead and hardware accuracy, energy efficiency of transceivers or MTC device lifetime related metrics, the receiver complexity which is of particular interest and which will be traded off against potential performance gains.

## VI. Concluding remarks

The expected outcome of 5GNOW is a much more robust transmission technique which efficiently exploits the huge number of (design) degrees of freedom available in a heterogeneous network and inherently supports a future differentiated service architecture ranging from MTC traffic to e.g. super-high rate video streaming etc. The S/T results of 5GNOW will also provide a fundamental understanding in all its aspects of the potential benefit of a non-orthogonal, asynchronous air interface. Altogether, the results will be condensed to a single proposal for a new non-orthogonal, asynchronous architecture for 5G mobile communication system.